\documentclass[preprint]{revtex4}

\usepackage{graphics}
\usepackage{bm}

\newcommand{\msolar}{\ensuremath{M_{\odot}}}
\newcommand{\beq}{\begin{equation}}
\newcommand{\eeq}{\end{equation}}
\newcommand{\yep}{ \left ( \frac{Y_e}{1-X_n} \right ) }
\newcommand{\yeplong}{(Y_e/(1-X_n))}

\begin{document}

 \title{Stellar Collapse Dynamics With Neutrino Flavor Changing Neutral Currents}

\author{Philip S. Amanik and George M. Fuller}
\affiliation{Department of Physics, University of California, San Diego, La Jolla, CA 92093-0319}

\date{\today}

\begin{abstract}
We perform one-zone simulations of the infall epoch of a pre-supernova stellar core in the presence of neutrino flavor changing scattering interactions.   Our calculations give a self-consistent assessment of the relationship between flavor changing rates and the reduction in electron fraction and re-distribution of initial electron lepton number among the neutrino flavors.  We discuss and include in our calculations sub-nuclear density medium corrections for flavor changing scattering coherence factors. We find that flavor changing couplings $\epsilon > 3\times10^{-4}$ in either the 
$\nu_e\leftrightarrow\nu_{\mu}$ or $\nu_e\leftrightarrow\nu_{\tau}$ channels result in a dynamically significant reduction in core electron fraction relatively soon after neutrino trapping and well before the core reaches nuclear matter density.  
\end{abstract} 


\maketitle

\section{Introduction}
Core collapse supernovae are exquisitely sensitive to lepton number violating processes.  This is because the infall (collapse) epoch of the pre-supernova core is characterized by low entropy\cite{bbal} and large lepton (electron and electron neutrino) degeneracy.  Nearly all of the pressure support stems from these degenerate leptons.
The effects of including neutrino flavor changing neutral current (FCNC) interactions in the infall stage of a core collapse supernova have recently been investigated in 
Ref. \cite{afg}.  It was noted there that neutrinos in the core of a collapsing star could undergo large numbers of scatterings due to the coherent amplification of the neutrino-quark flavor changing neutral current cross section for elastic scattering on heavy nuclei.   Such interactions could cause significant numbers of electron neutrinos in the core to be converted to mu and tau neutrinos.   In turn, this would open phase space for further electron capture and thereby significantly impact the pressure, homologous core mass, and the initial shock energy.

The explosion of core collapse (Type II, Ib, and Ic) supernovae is believed to be the result of gravitational collapse, subsequent hydrodynamic bounce of the star's core, and release of gravitational binding energy into neutrinos which ultimately provide the energy to revive and sustain the shock \cite{c&w,wilson,b&w,bethe,b&y,m&b,jkr}.
One important feature of the model is that the entropy of the core is low ($s/k\sim 1$) and nucleons remain bound in nuclei during most of the collapse.  
The number of electrons in the core (hence, the pressure and homologous core mass) is governed by the electron capture reaction $e^- + p \leftrightarrow \nu_e + n$.  When the neutrino mean free path becomes smaller than the size of the core (because of scattering on heavy nuclei) the neutrinos become trapped.  They thermalize quickly and comprise a degenerate Fermi-Dirac sea.  When the $\nu_e$ 
Fermi level becomes high enough, electron capture is blocked and \emph{net} reduction in $Y_e$ (where $Y_f\equiv (n_f-n_{\bar{f}})/n_b$) 
no longer occurs on dynamical time scales.  However, re-distribution of electron lepton number between $\nu_e$'s and electrons will still occur as the density rises and the nuclear composition changes. 

Any further changes in the core's electron fraction during the collapse could result in a change in the collapse dynamics and explosion mechanism\cite{hix}.  
Including neutrino FCNC interactions in the collapse model causes greater  reduction in $Y_e$ during infall.  This is because when electron neutrinos change flavor by scattering, holes open in the $\nu_e$ sea and the electron capture reaction can procede.

Neutrino-quark FCNCs of the form 
\begin{equation}
\mathcal{L}=\frac{G_F}{\sqrt 2}\bar{\nu}^j\gamma^\mu\nu^i\bar{q}\gamma_\mu(\epsilon^q_{V_{ij}}+\epsilon^q_{A_{ij}}\gamma^5)q
\label{fcnclag}
\end{equation}
were considered in Ref. \cite{afg}. Here, the parameters $\epsilon^q_{V_{ij}}$ and $\epsilon^q_{A_{ij}}$ quantify the strength of the FCNC relative to the Fermi constant $G_F$.  Current experimental constraints\cite{davidson} on the FCNC couplings are $\epsilon^q_{V_{e\mu}} < 10^{-3}$ for the channel $\nu_e\leftrightarrow\nu_\mu$ and $\epsilon^q_{V_{e\tau}} < 5\times 10^{-1}$
 for the channel $\nu_e\leftrightarrow\nu_\tau$. (Similar, and in some cases better, constraints on these interactions may be possible from solar and atmospheric neutrinos\cite{f&l}.)
 
The cross section for neutrino flavor changing elastic scattering on heavy nuclei, mediated by the FCNCs of Eq. (\ref{fcnclag}), was calculated in Ref. \cite{afg} and a coherent amplification was found. Using this cross section, and employing values of the coupling constant up to and beyond current experimental constraints, Ref. \cite{afg} gave estimates for the number of neutrino flavor changing scattering events which could occur in the core.  The resulting reduction in $Y_e$ and implications for the stellar collapse model were then discussed in a qualitative sense.

In this paper we present results of a one-zone calculation of the infall epoch of a pre-supernova star with neutrino-quark FCNCs included. Our code gives a more accurate accounting of scattering rates and the change in $Y_e$ than do the estimates of Ref. \cite{afg} and we are able to account for some of the  feedback in the system.  We model neutrino scattering with nuclei in the core medium and account for sub-nuclear matter density structure effects.  By contrast, Ref. \cite{afg} employed neutrino-nucleus vacuum cross sections with no accounting for medium effects.  Reference \cite{afg} estimated which values of $\epsilon$ would give a fast enough FCNC scattering rate such that reduction in $Y_e$ would be possible. Here we actually compute what the reduction in $Y_e$ is for various values of $\epsilon$, including values below the best experimental bounds. We have discovered that maximal reduction in $Y_e$ is possible for values of $\epsilon$ smaller than the best experimental bound in the $\nu_e\leftrightarrow\nu_{\tau}$ channel, and that dynamically significant reduction in $Y_e$ is possible for values of $\epsilon$ smaller than the best experimental bound in the $\nu_e\leftrightarrow\nu_{\mu}$ channel. In section II we describe our code and method of computing the change in electron fraction.  In section III we discuss our results and their meaning for the stellar collapse model.  In section IV we list the key approximations in our calculation and give an assessment of the possible impact of the potential uncertainties introduced by these. In section V we give conclusions.

\section{One-Zone Core Collapse Simulation} 

We seek a self-consistent relationship between FCNC rates and the possible reduction in core electron fraction resulting from these processes.    We  simulate the core collapse with a one-zone calculation which computes reactions rates (including FCNC rates), thermodynamic quantities, equation of state (EOS) quantities, and electron and neutrino fractions.  Though one-zone calculations obviously do not include a sophisticated treatment of hydrodynamics or neutrino transport, and can contain many assumptions, they have been used successfully to model feedback between weak interactions and nuclear equation of state parameters in the infall epoch of stellar collapse \cite{bbal,f82}.  The validity they have rests on three key and non-controversial features of the infall epoch: low entropy; $e^-$ and 
$\nu_e$ degenerate conditions; and lepton capture rates dominated by the energetics scales associated with the high lepton Fermi levels.  Note that these key features are also confirmed by large sophisticated numerical simulations \cite{b&y,m&b,jkr}.
 
\subsection{Description of Calculation}
The code is a modified  version of that used in Ref. \cite{f82}.   In the calculations done here and in Ref. \cite{f82}, a single zone (with initial electron fraction $Y_e$, density $\rho$, temperature $T$, entropy per baryon $S$, neutron mass fraction $X_n$, neutron kinetic chemical potential $\mu_n$, and neutron-proton kinetic chemical potential difference $\hat{\mu}$) is evolved assuming a uniform collapse rate.  A standard Newton-Raphson algorithm is employed.  As the density increases, the electron Fermi energy rises and the electron capture rate increases.  At each density step the electron capture rate and collapse rate are used to find the change in electron fraction $\Delta Y_e$, and then $\Delta Y_e$ is used to estimate a change in entropy $\Delta S$.  The updated values for $\rho$, $Y_e$, and $S$, along with explicit expressions for $S$ and $X_n$ are used by a routine which increments the temperature and utilizes a two-dimensional root finder to iteratively compute $X_n$ and $T$.  The mean nuclear mass $A$, $\mu_n$, and $\hat{\mu}$ are also found during this iterative process.    The EOS formulae used  \cite{bbal,f82} for the mean nuclear mass, nucleon chemical potentials and nucleon-to-baryon ratios are based on a finite temperature liquid drop model (see Ref.s \cite{bbal,bbp}) with a representative mean heavy nucleus and a sea of dripped neutrons.  This is discussed in Appendix \ref{codeap}. 

The core's electron fraction changes because of electron capture reactions.  After neutrinos have become trapped in the core and the $\nu_e$'s build up a degenerate Fermi sea, an equilibrium situation obtains: $Y_e$ and the net number of $\nu_e$'s per baryon 
$Y_{\nu_e}$ no longer change appreciably even though electron capture reactions, and the inverse reactions, are taking place.  In the presence of FCNCs,  $\nu_e$'s change flavor. As a result, phase space is opened
allowing \emph{net} electron capture to occur and causing further reduction in $Y_e$ and in overall electron lepton number. 

The physical reason for the reduction of $Y_e$ is that electron capture reactions lower the number of electrons in the core. We count the reduction of $Y_e$ in two ways.  The first way uses the electron capture rate for reactions occurring before equilibrium is established.  The second way counts electron captures which occur as a result of  phase space opening in the $\nu_e$ sea, secondary to flavor changing scattering events. This is computed using the  neutrino flavor changing rate.  Counting the reduction in $Y_e$ in the first way applies only until complete beta-equilibrium obtains, while counting in the second way applies both before and after beta-equilbrium is established.  We will discuss the first way here, and discuss the second way in Subsection \ref{Dye}, after we present the neutrino flavor changing scattering rate.  

The rates of electron capture on free protons and heavy nuclei are derived in Ref. \cite{f82} and denoted respectively by $\lambda_{\rm fp}$ and $ \lambda_{\rm H}$.
The total rate of electron capture per baryon is 
\beq
\frac{d Y_e}{dt} = - X_p \lambda_{\rm fp}  - \frac{X_H}{A} \lambda_{\rm H}, \label{yedot}
\eeq
where $X_H\approx (1- X_n -X_p)$ is the mass fraction of heavy nuclei and $X_n$ and 
$X_p$ are the neutron and proton mass fractions, respectively. (The number abundance of heavy nuclei relative to baryons is $Y_H=X_H /A $ while the corresponding abundances of the free nucleons are $Y_n=X_n$ and $Y_p=X_p$.) We take the alpha particle mass fraction to be negligible, consistent with the low entropy infall conditions.
Combining Eq. (\ref{collapserate}) for the collapse rate and Eq. (\ref{yedot}) we have
\beq
\frac{dY_e}{d\rho} = \left [ - X_p \lambda_{\rm fp}  - \frac{1 - X_n - X_p}{A} \lambda_{\rm H} \right ] \left ( \frac{10^{-12}}{\rho_{10}^{3/2}}\right ) \frac{\rm s}{{\rm g}/{\rm cm}^3}
\label{dyedrho}
\eeq
where $\rho_{10}=\rho /(10^{10} {\rm g}/{\rm cm}^3)$.
This gives $\Delta Y_e$ at each density step and is used to find $Y_e$ until the beta equilibrium condition is imposed.

At the onset of collapse, electron neutrinos created from electron capture stream freely out of the core.  As nuclei become more neutron-rich and the cross section for ordinary coherent neutral current scattering becomes appreciable, high energy neutrinos begin to be trapped in the core and start to equilibrate.  This occurs for a matter density of $\rho \sim 10^{12}{\rm g}/{\rm cm}^3$.  We start neutrino trapping at the density $\rho_{\rm trap}=5\times 10^{11}{\rm g}/{\rm cm}^3$.  When the simulation reaches  density $\rho_{\rm trap}$ it begins accounting for the $\nu_e$'s getting trapped in the core. The $\nu_e$ fraction $Y_{\nu_e}$ is found  by calculating $\Delta Y_e$ with equation (\ref{dyedrho}) at each density step and imposing the condition $\Delta Y_{\nu_e}= \Delta Y_e$.
This of course is an approximation because it means that after the ``trapping density'' is reached, every electron capture creates a neutrino which becomes trapped.
In reality, neutrinos are not trapped instantaneously at some density, but rather are gradually trapped as neutrino diffusion times increase.  Also, $\nu_e$ cross sections scale as neutrino energy squared so some lower energy neutrinos will still be escaping after the higher energy neutrinos have become trapped. Though our model for neutrino trapping is obviously simplistic, it gives values for $Y_{\nu_e}$ consistent with the currently accepted core collapse model \cite{c&w,wilson,b&w,bethe,b&y,m&b,jkr}. 

One-zone collapse calculation results for various cases are shown in Tables 
\ref{eps0tab}-\ref{eps1-2tab}. These give density $\rho_{10}$, electron fraction $Y_e$, mu plus tau neutrino fraction $Y_{\nu_\mu}+Y_{\nu_\tau}$, temperature $T$ in MeV, entropy per baryon $s/k$ in units of Boltzmann's constant, free neutron fraction $X_n$, neutron kinetic chemical potential $\mu_n$ in MeV, mean nuclear mass $A$, mean nuclear radius $r_{\rm nuc}$ in fm, separation distance between nuclei $D_{\rm sep}$ in fm, average neutrino energy $E_{\nu}$ in MeV, and average neutrino de Broglie wavelength $\lambda_{\nu}$ in fm. Note that our crude neutrino trapping and neutrino sea filling schemes give some unphysical results. For example, once neutrino trapping is enforced at $\rho_{\rm trap}=5\times10^{11}{\rm g}/{\rm cm}^3$, the entropy is taken as constant. However, the system is still out of chemical equilibrium so increments in density give a temporary and small drop in temperature. This gives a negligible overestimate of the FCNC effects near trapping because it produces slightly larger nuclei, with of order a few extra nucleons. Likewise, because we do not consider $\nu_e$-capture self consistently with electron capture, and because our liquid drop equation of state is inaccurate for high density and high neutron excess, runs with values $\epsilon \gtrsim 10^{-3}$ acquire a positive neutron kinetic chemical potential at sub-nuclear but large densities.  Note however, that where this happens FCNCs have already had a significant effect.

When the electron neutrino fraction reaches $Y_{\nu_e}=0.05$, we impose the condition that beta-equilibrium has been reached.  We have chosen $Y_{\nu_e}=0.05$ as final equilibrium value of $Y_{\nu_e}$ to be consistent with the currently accepted supernova model and large scale numerical simulations 
\cite{c&w,wilson,b&w,bethe,b&y,m&b,jkr}.  At this point, in the absence of neutrino FCNCs and density- and composition-driven equilibrium shifts, the values for $Y_e$ and $Y_{\nu_e}$ are the final values. This is another approximation since equilibrium does not actually obtain instantaneously.  In Table \ref{eps0tab} we have given results for a run of our simulation with flavor changing interactions turned off.  The run used to produce this table started from a density 
of $3.7\times 10^{9} {\rm g}/{\rm cm}^3$ and went to a final density of  
$3.8\times 10^{13} {\rm g}/{\rm cm}^3$.  This table shows the changes in $Y_e$ and $Y_{\nu_e}$.  As soon as the simulation starts, $Y_e$ is decreasing. After the trapping density, $Y_{\nu_e}$ starts to increase and once it reaches 0.05, $Y_e$ and 
$Y_{\nu_e}$ no longer change.

\subsection{Neutrino Flavor Changing Rate in the Core}
Understanding the nuclear composition and equation of state in the core of a collapsing star is an active area of research. It is believed that as the core approaches nuclear matter density, $\rho\sim 10^{14} {\rm g}/{\rm cm}^3$, the nuclear component undergoes a series of phase transitions as the individual nuclei merge and, in fact, eventually cease to exist \cite{ls}.  During these phases, i.e. ``pasta phases,'' the nuclear matter may take the form of rods, sheets or tubes.    Recent work has focused on how neutrinos scatter coherently on stuctures in these phases. (See for example Ref. \cite{horowitz}.)   As outlined above, we use a liquid drop model to describe the nuclear component in the core.  We use this model to describe the core only up to a density of $\rho = 3.8\times 10^{13} {\rm g}/{\rm cm}^3$.  The liquid drop model may not be valid over the whole density range where we have used it.  By only running our simulation up to a maximum density which is an order of magnitude below nuclear density, we avoid most of the density range where it is guaranteed to be inaccurate.  

In Tables \ref{eps0tab}-\ref{eps1-2tab} we show values for the mean nuclear mass $A$, radius of the mean nucleus $r_{\rm sep}$, and the nuclear separation distance, $D_{\rm sep}$. It should be kept in mind that a nuclear statistical equilibrium mix of nuclear sizes and masses will exist in the core.  The mean nuclear mass is taken from Eq. (\ref{nucmass}) and the nuclear radius is found from 
$r_{\rm nuc} \approx A^{1/3}\,$fm. To calculate the separation distance between nuclei, we assume each nucleus is in a Wigner-Seitz cell with cell volume $V_c=1/n_H$, where $n_H=\rho N_A Y_H$ is the number density of heavy nuclei and $N_A$ is Avagadro's number. Then $D_{\rm sep}= 2 R_c$, where $R_c$ is the radius of the cell. 

In Table \ref{eps0tab}, for example, we see that values for mean nuclear mass become as high as $A\approx 300$.  At earlier epochs and lower densities in our simulation, we see that the nuclear masses are below $A=200$. These values are consistant with Ref. \cite{ls}, which in turn, is a foundation for modern full-scale supernova simulations\cite{a-b}. At the higher densities, the nuclei have radii as large as 7 fm, and are separated by at least 37 fm.  In the middle of the density range, the nuclei have radii less than 5 fm and are separated by at least 40 fm and as much as 80 fm.  According to our liquid drop model, the values for $A$, $r_{\rm nuc}$, and $D_{\rm sep}$ indicate that we are not close to densities where the nuclei merge.  Therefore, up to a density of $3.8\times 10^{13} {\rm g}/{\rm cm}^3$, it is reasonable to consider coherent scattering of neutrinos on individual nuclei in the core. However, in reality one should consider the more complicated problem of neutrino coherent scattering on the three dimensional structures in the pasta phases \cite{horowitz}. 

We calculate the rates for neutrino flavor changing scattering on free nucleons and on the mean nucleus.  Coherent scattering on nuclei is the dominant flavor changing reaction.  In the core it can be necessary to take account of scattering interference effects arising from conditions where more than one nucleus resides within a neutrino DeBroglie wavelength. This is an issue whenever $1/E_\nu \gtrsim R_c$ \cite{sanjay}, where $E_\nu$ is the average neutrino energy. This condition means that the neutrino DeBroglie wavelength is comparable to or larger than the distance between nuclei.  In a relativistically degenerate Fermi gas the average neutrino energy is 3/4 of the neutrino chemical potential $\mu_{\nu_e}$, where $\mu_{\nu_e} \approx 11.1\,{\rm MeV} (2 \rho_{10} Y_{\nu_e})^{1/3} $.  In Tables \ref{eps0tab}-\ref{eps1-2tab} we show average neutrino energies and average neutrino DeBroglie wavelengths, $\lambda_\nu$.  Recalling that $R_c=D_{\rm sep}/2$, and comparing $R_c$ to $\lambda_\nu$, we see that we are in a regime where interference will occur.  In Appendix \ref{corap} we present the cross section for neutrino scattering with nuclei in a medium.

The neutrino flavor changing scattering rate employed here is as follows.  First, the electron neutrino flavor changing scattering rate per mean nucleus is 
\begin{equation}
\lambda_A =  \left ( \rho N_A Y_{\nu_e} \right ) c
\frac{\epsilon^2 G_F^2}{\pi}  \mathcal{I} (2N+Z)^2 E_\nu^2.
\end{equation}
Here, $\rho$ is the matter density, $c$ is the speed of light and $\mathcal{I}$ is a factor which corrects for interference.
This rate was obtained by multiplying the $\nu_e$ flux by the cross section given in Eq. (\ref{cs}).  The coherent amplification factor is $(2N+Z)^2$, where $N$ and $Z$ are the number of neutrons and protons in the mean nucleus.  We have dropped sub and superscripts on the FCNC coupling $\epsilon$ so that this rate is generic and refers to scattering on a $d$-quark in either the $\nu_e\leftrightarrow\nu_\mu$ or $\nu_e\leftrightarrow\nu_\tau$ channel.
The rate for electron neutrino flavor changing scattering on free nucleons is
\begin{equation}
\lambda_{\rm fn} = \left ( \rho N_A Y_{\nu_e} \right )c \frac{\epsilon^2 G_F^2}{\pi} 2 
E_\nu^2. \label{ratefn}
\end{equation}
We do not concern ourselves with accounting for the fact that the free nucleons are in a medium; the cross section for coherent scattering on nuclei is larger by a few orders of magnitude, so the nuclei dominate the FCNC opacity.
The total neutrino flavor changing rate per baryon is
\begin{eqnarray}
\lambda_{\rm b} &=& \sum_i Y_i \lambda_i \nonumber \\
&=& \lambda_{\rm fn}\left\{ 1-X_H +
\frac{X_H}{A} \frac{\mathcal{I}}{2} 
\left [A\left(2-\frac{Y_e}{1-X_n}\right) \right]^2 \right \}. \label{rateb}
\end{eqnarray}

\subsection{Counting the Reduction in Electron Fraction} \label{Dye}
Given that the parameter $\epsilon$ is currently constrained by experiment to be smaller than $\sim10^{-1}$ ($\nu_e\leftrightarrow\nu_\tau$) or $\sim10^{-3}$ ($\nu_e\leftrightarrow\nu_\mu$), the flavor changing scattering rates could be less than the  electron capture rate.  We work in a limit such that 
whenever an electron neutrino undergoes a flavor changing scattering and opens a hole in the $\nu_e$ sea, the hole is immediately filled by a $\nu_e$ produced via an electron capture. We argue later in this section that our calculated scattering rates justify this approximation. In this limit, FCNC transformation of electron neutrinos into  mu and tau neutrinos will not change $Y_{\nu_e}$. Rather, the change in electron fraction $\Delta Y_e$ is simply minus the total change in the sum of mu and tau neutrino fractions
\begin{equation}
\Delta Y_e = -\Delta (Y_{\nu_\mu}+ Y_{\nu_\tau}). \label{delye}
\end{equation}
In other words, the net reduction in $Y_e$ equals the sum of the net increase of the mu and tau neutrino fraction. 
The change in $Y_{\nu_\mu}+Y_{\nu_\tau}$ in a density step is 
\begin{equation}
\Delta (Y_{\nu_\mu}+Y_{\nu_\tau}) = \Delta \rho \lambda_{\rm b} \frac{10^{-12}}{\rho_{10}^{3/2}}\left ( \frac{\rm s}{{\rm g}/{\rm cm}^3}\right), \label{delmutau}
\end{equation} 
where  $\rho_{10}=\rho/(10^{10} {\rm g}/ {\rm cm}^3)$ and $\lambda_b$ is given in 
Eq. (\ref{rateb}). (This expression is analogous to Eq. \ref{dyedrho}.)
The total reduction in $Y_e$ stemming from neutrino FCNCs is found from Eq.s (\ref{delye}) and (\ref{delmutau}) by summing the increments from each density step. We impose the 
condition that the FCNC interactions are turned on in our one-zone simulation at the neutrino trapping density, which we take to be $\rho_{trap}=5\times 10^{11} {\rm g} / {\rm cm}^3$, a value consistent with large scale simulations.  Before trapping, neutrinos are freely streaming out of the core and net electron capture is not yet blocked.   It would not matter to the core's final value of $Y_e$ if electron neutrinos changed flavor before streaming out of the core.  

\begin{figure}
\resizebox{9cm}{!}{\includegraphics{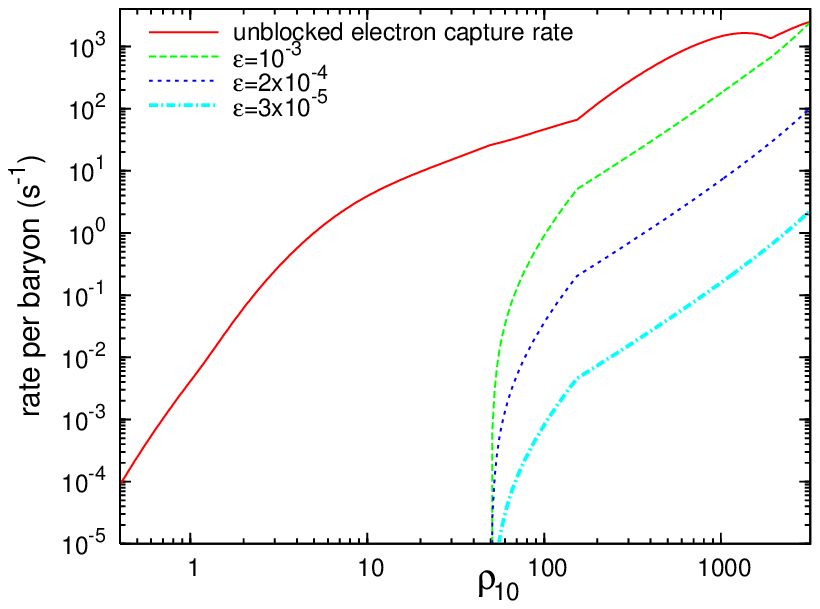}}
\caption{(color online). Neutrino FCNC scattering rates as a function of density for the indicated values of FCNC coupling $\epsilon$. The solid curve shows the unblocked electron capture reaction rate.}
\label{rates}
\end{figure}

In Figure \ref{rates} we have plotted the reaction rates from our calculations as a function of core density.  In this log-log plot, rates are given as number of reactions per baryon per second, and the core density is given in terms of $\rho_{10}$.  As mentioned above, FCNC interactions are started in the simulation at a density of $\rho_{10}=50$.  The dotted curve in the plot shows the unblocked electron capture reaction rate, i.e. the reaction rate computed as if there was no blocking of the final state $\nu_e$. This should not be confused with the actual net rate of electron capture (neutronization rate) in the core which is, of course, affected by blocking and by the reverse $\nu_e$ capture reaction.  We show the unblocked reaction rate to illustrate the relative size of the FCNC rates.  Note that the FCNC rates for values of $\epsilon\leq 10^{-3}$ are smaller than the electron capture rate. Therefore, for values of $\epsilon\leq 10^{-3}$, our assumption for computing $\Delta Y_e$ (where an electron capture is assumed to occur immediately when an electron neutrino changes flavor) is definitely justified.  We expect that the assumption is also valid for values of $10^{-3}\leq\epsilon\leq 10^{-1}$, though this range of epsilon is more complicated.  For example, $\nu_{\mu,\tau}$ neutrinos can undergo flavor changing scattering into $\nu_e$'s and sometimes fill holes before electron captures can. This scenario is discussed further in Section \ref{dyefcnc}.  This range in $\epsilon$ would be best modeled by a more detailed simulation which calculates neutrino transport. We anticipate that such a simulation would very our assumption.  Finally, for values of $\epsilon > 10^{-1}$, the FCNC cross section is comparable to the electron capture cross section and our assumption definitely does not hold.  We therefore do not include these values of $\epsilon$ in our simulation.

From Figure \ref{rates} we see that even though the FCNCs are weaker and slower than Standard Model weak interactions, flavor changing scattering can nevertheless be significant.  For $\epsilon=10^{-3}$, we see that at a density of $\rho_{10}=100$ there are $\sim 10$ flavor changing scatterings per baryon per second.  At this density there are $\sim 10^{36}$ baryons per cubic centimeter in the core.  Clearly then, there are a large number of electron neutrinos changing flavor as the core passes through this density.  Of course, even greater numbers of neutrinos will change flavor as the collapse proceeds to higher densities.
 
\section{FCNC-Induced Reduction in Electron Fraction and Alterations in Core Physics} \label{dyefcnc}

\begin{figure}
\resizebox{9cm}{!}{\includegraphics{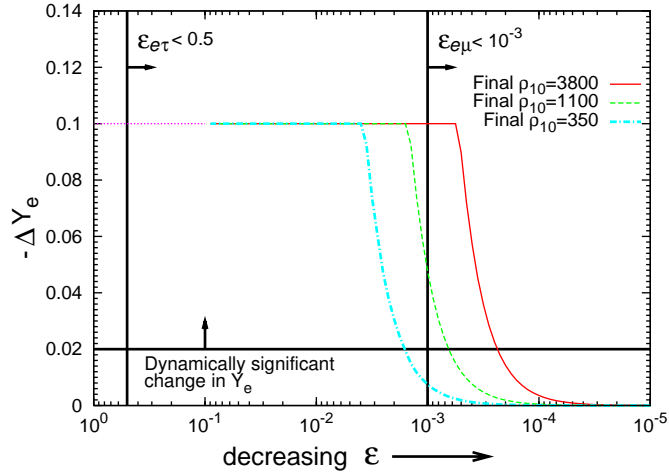}}
\caption{(color online). Magnitude of the net decrease in electron fraction $(-\Delta Y_e)$ as a function of $\epsilon$, for collapse up to various final densities. Curves become dotted for $\epsilon > 10^{-1}$ because we have not included this range of $\epsilon$ in our calculation; the dotted line is included for ease of reading the figure. The horizontal line indicates a threshold change in electron fraction ($-\Delta Y_e > 0.02$) beyond which significant alteration in core physics can be expected. The two vertical lines give the current experimental bounds on $\epsilon = \epsilon_{e\tau}$ (for $\nu_e\leftrightarrow\nu_\tau$) and  
$\epsilon = \epsilon_{e\mu}$ (for $\nu_e\leftrightarrow\nu_\mu$).}  
\label{epcon}
\end{figure}

Considering the FCNC rates in the core is useful in demonstrating that there are indeed a large number of neutrinos changing flavor, but the quantity which can be most important for the dynamics of the supernova model is the electron fraction.  We have calculated the core's total $\Delta Y_e$ stemming from FCNCs as a function of the FCNC coupling constant 
$\epsilon$.  As mentioned in the previous section, we find the total $\Delta Y_e$ for each collapse simulation by summing Eq. (\ref{delmutau}) from each density step.  
The results are plotted in Fig. \ref{epcon}.
Curves are shown for three different simulations, up to final densities of $\rho_{10}=350$, $\rho_{10}=1140$ and $\rho_{10}=3800$.  The range of the continuous parameter $\epsilon$ is $10^{-1}$ to $10^{-5}$. The figure has vertical lines which show current values of experimental constraint on the epsilon parameter.  The dotted line from $\epsilon=1$ to $\epsilon=10^{-1}$ is included for ease in interpreting the figure, but we did not include these $\epsilon$ values in our calculation. These values of epsilon are not strictly covered by the limiting case of $e^-$ capture rates being faster than the $\nu_e$ flavor changing rate.  We discuss the case of large values of $\epsilon$ below. 

In our simulation we assume a maximum trapped neutrino fraction of $Y_\nu=0.05$ for each of the three flavors.  Therefore, by Eq. (\ref{delye}), the maximum reduction possible for the electron fraction is $\Delta Y_e = - 0.1$.  Of course, if one were to consider a different value for the maximum trapped neutrino fraction, the maximum $\Delta Y_e$ would be different.  As we will discuss below, and as marked on the figure, a dynamically significant reduction in $Y_e$ can be as low as $\Delta Y_e = - 0.02$ \cite{hix}. When the simulation
runs to $\rho_{10}=3800$, our results show that the maximum reduction in $Y_e$ occurs even for values of $\epsilon$ as low as $\epsilon=10^{-3}$. To put this in context, 
$\epsilon=10^{-3}$ is coincident with the best current experimental constraint on the $\nu_e\leftrightarrow \nu_\mu$ channel and is orders of magnitude smaller than the best current experimental constraint on $\nu_e\leftrightarrow\nu_\tau$.  For couplings 
$\epsilon <10^{-4}$, we see that reduction in $Y_e$ due to FCNCs ceases to be significant.  
 
Figure \ref{epcon} also has curves produced from simulations which were run to lower final densities.  A reason for restricting our calculations to lower final densities is that our schematic liquid drop model equation of state is more reliable at lower density.  However, since in these cases 	the FCNCs are not active for as long, there is less time for reduction in $Y_e$ to accumulate.  The figure shows that even if FCNCs are active for only a short duration after trapping, significant and/or maximal reduction of $Y_e$ can occur for values of $\epsilon$ allowed by current experimental bounds.  

In Tables \ref{eps1-3tab} and \ref{eps1-2tab} we show data from the simulation with active FCNCs. These tables can be compared to Table \ref{eps0tab} which comes from a simulation without FCNC interactions. Table \ref{eps1-3tab} and Table \ref{eps0tab} both go to the same final density of $\rho_{10}=3800$. As can be seen in 
Table \ref{eps1-3tab}, $Y_{\nu_\mu} + Y_{\nu_\tau}$ reaches the maximum level (and maximum reduction in $Y_e$ is obtained) at a density of 
$\rho_{10} =  2.87\times 10^3$. At this density the mean nucleus has a mass of 
$A=334$.  For the simulation without FCNCs included, the mean nucleus has a mass of $A=292$ at this same density.  The mean nucleus is larger in the simulation with FCNCs because the increased electron capture has caused nuclei to be more neutron rich.  For the larger value of $\epsilon=10^{-2}$, Table 
\ref{eps1-2tab} shows that maximum reduction in $Y_e$ occurs already at a density of 
$\rho_{10}\approx 200$. At this density the mean nucleus has a less exotic size,  $A\approx 123$.

For our scenario where electron capture is fast compared to FCNCs, whenever an electron neutrino changes flavor, the hole in the $\nu_e$ sea is assumed to be filled immediately by a $\nu_e$ produced via electron capture.  We have ignored the possibility of mu or tau neutrinos undergoing FCNC scattering and changing into electron neutrinos, thus filling holes before electron capture can occur.  It is most likely that holes would be filled by $\nu_e$'s produced by electron capture. If a situation arises in the core where some holes are  filled by mu or tau neutrinos that changed to electron neutrinos, the number of $\nu_e$'s in the core would still remain the same. The $\nu_e$ fraction still remains fixed at its maximum value of $Y_{\nu_e}=0.05$ and Eq. (\ref{delye}) remains valid.  In such a situation it is still possible for maximal reduction of $Y_e$ to occur.   

Situations other than those covered by our limiting case are also possible. These scenarios could occur for values of $\epsilon$ close to $10^{-1}$.  One example is if large FCNC rates cause significant numbers of mu and tau neutrinos  to change flavor and seriously compete with electron capture in filling holes in the $\nu_e$ sea.  In this case, $Y_{\nu_e}$ still remains fixed at its maximum value and Eq. (\ref{delye}) is again valid, but maximal reduction in $Y_e$ would \emph{not} occur.  Another example is if the  FCNC rates are greater than the electron capture rate.  In this case we expect that the neutrino seas would equilibrate and all reach the same level, lowering the Fermi level of  the $\nu_e$ sea before electron capture reactions can replenish it.  Instead of remaining fixed, $Y_{\nu_e}$ would be lowered initially. Maximum reduction in electron fraction may or may not occur.  Equation (\ref{delye}) is not valid in this situation and the lepton number distribution in the core would be different from that of our limiting case. Even if maximum possible reduction of $Y_e$ does not occur during infall, we stress again that it only takes $\Delta Y_e \approx - 0.02$ to produce a significant alteration in core physics \cite{hix}.  

According to the current model\cite{c&w,wilson,b&w,bethe,b&y,m&b,jkr} for core collapse supernovae, reduction of the core's electron fraction during infall will hinder the supernova explosion\cite{bbal,hix,f82}.  Electrons in the core influence the collapse dynamics through the degeneracy pressure they provide.  In particular, the number of electrons determines the size of the homologous core, $M_{\rm hc} \approx 5.8 Y_e^2 \msolar$.  A lower electron fraction, and consequentially smaller homologous core, hinders the explosion in two ways.  If the inner core is smaller, there is more material in the outer region of the core for the shock to photodissociate before reaching the outer envelope of the star.  Therefore, the shock has less energy available to eject the outer envelope and cause the explosion.  A smaller inner core also has a smaller gravitational potential and so the outer core material has lower infall kinetic energy.  The infall energy gets converted to the initial outgoing energy of the shock wave at bounce. (The initial shock energy scales as $Y_e^{10/3}$ as shown in Ref. \cite{f82}.) Therefore, a smaller inner core results in a weaker shock.  A weaker shock and greater loss of energy for the shock during its progression through the outer core may make an explosion more difficult to obtain.
 
We have seen that including FCNCs in the supernova model causes $Y_e$ to be lowered and thus disfavors a successful explosion, or at least, can significantly alter the model.  The FCNCs change the core's lepton number content.  A standard collapse model would suggest that at bounce there would be a net electron lepton number in the core, but no net mu or tau lepton number. By contrast, with FCNCs there could arise significant net mu and tau lepton numbers resident in seas of $\nu_\mu$'s and $\nu_\tau$'s. Of course, in this case we would still have sizeable electron lepton number residing in the electrons and the $\nu_e$'s. This might have an interesting effect on the expected supernova neutrino signal, as speculated in Ref. \cite{afg}. Neutrino medium-enhanced flavor mixing (oscillations) above the neutron star will occur and will affect the signal \cite{fmmw,fhm,p&r,b&f,bal&yuk,hrs,dfq,dfcq1,dfcq2,f&q}. Since neutrino mass-squared differences are known and mixing parameters (i.e., $\theta_{13}$) may be better constrained in the future, it may be possible to predict the effects of flavor mixing and, upon detection of a supernova signal, subtract these out to identify signatures of FCNCs. An FCNC-engendered excess of $\nu_\mu$'s and/or $\nu_\tau$'s might also result in altered neutrino energy/entropy transport in the proto-neutron star.

\section{Detailed Summary and Discussion of Approximations}
We now summarize the approximations we have made and compare our calculation with modern detailed simulations.  List of approximations:
\begin{itemize}
\item One-zone
\item Constant collapse rate
\item Fermi Dirac distributions for neutrinos
\item Instantaneous trapping
\item Liquid drop model with mean representative nucleus
\item vacuum cross sections for neutrino nucleus interactions
\item ignoring medium effects and nuclear collective modes
\item electron capture rate is faster than flavor changing rate
\end{itemize}
We first point out that even though our calculation has approximations, it is based on the fact \cite{bbal} that during the infall stage of collapse the core is a low entropy environment. Our equation of state is simple and based on Ref. \cite{bbal}. The Ref. \cite{bbal} EOS captures the essential features of more sophisticated treatments, for example Ref. \cite{ls}.  Moreover, there is no difference in the \emph{underlying} physical principals upon which our calculation and the detailed numerical simulations are based.

Large detailed simulations follow the profile of the whole core, rather than a single zone as we have done. They calculate neutrino distributions and transport rather than imposing and modifying the distributions by hand.  Neutrino trapping occurs ``naturally'' and the rate of the collapse is calculated self consistently within these simulations.  In our case, we have not calculated these parameters in the presence of FCNCs.  This is an approximation since we expect the physical effect of $\nu_e\rightarrow\nu_{\mu,\tau}$ will give some feedback on the system and these parameters will be modified.  We point out however that the values we imposed (for example $Y_e$ in the range 0.3-0.35, and collapse rate as in Eq.~(\ref{collapserate}) are consistent with modern simulations.  

As for nuclear matter, large detailed simulations treat this in the same way as we have.  Nuclear matter at high density and temperature is still a wide area of research.  Representing the nuclear matter in the core by an average nucleus, using a vacuum cross section for this nucleus which is in the core, and ignoring any other medium effects is an approximation.  The large simulations make these same approximations when modeling the standard interactions.  We have accounted for one medium effect: multiple target scattering when the neutrino's wavelength is comparable to the nuclear separation distance. 

To summarize, we have made approximations involving the thermodynamic profile and the core's dynamical quantities, neutrino transport, and nuclear matter.   The thermodynamic and dynamical quantities we have used are based on the assumption of a low entropy core and are consistent with modern simulations.  Our treatment of nuclear matter is similar to that in large scale numerical simulations. Neutrino transport near the trapping point is the issue that could be most improved upon by modeling FCNCs in a full simulation with Boltzmann neutrino transport. Finally, we point out that if a full simulation were performed to model these interactions, our qualitative conclusions would not change. Furthermore, our quantitative results are conservative and we expect a full simulation would indicate even smaller values of epsilon for which flavor changing significantly impacts core physics. This is because we have stopped our calculation at modest densities but larger numerical simulations would be able model these ultra high density regimes.

\section{Conclusions}
We have used a one-zone core collapse simulation to investigate some effects of including neutrino flavor changing interactions in the supernova model.  We have calculated the reduction in $Y_e$ as a function of the coupling constant $\epsilon$ for collapse simulations that run up to density 
$\rho= 3.8\times 10^{13} {\rm g} / {\rm cm}^3$.  For values of the interaction coupling constant $\epsilon \gtrsim 5\times 10^{-4}$ in either the $\nu_e \leftrightarrow \nu_\mu$ or $\nu_e \leftrightarrow \nu_\tau$ channel we have found that maximal reduction in the core's electron fraction can occur.  (See Fig. \ref{epcon}.)

This work gives a more accurate and quantitative calculation of the effects of FCNCs than do the qualitative estimates given in Ref. \cite{afg}. Here, we are able to account for the FCNC rate's dependence on density, and the feedback on the rates as the core becomes more neutron-rich as a result of increased net electron capture.   However, a more accurate treatment of these interactions is possible and is warranted.  Some of our approximations were made for ease of calculation, while others were made to handle physics that is not yet well understood.  Even with our conservative treatment (e.g., not following FCNCs beyond a density $\rho = 3.8\times 10^{13} {\rm g} / {\rm cm}^3$), we were able to demonstrate how strong the effect of neutrino flavor changing interactions can be on the  infall epoch physics. At very high densities near core bounce, we expect FCNCs still to be appreciable and to continue to cause net reduction in $Y_e$.  If a full simulation was preformed which included FCNCs and ran all the way to bounce density, properly accounting for neutrino scattering with nuclear matter, our results lead us to believe that significant reduction in $Y_e$ would occur for values of $\epsilon$ even smaller than we have found here.

The EOS and neutrino scattering cross sections in nuclear matter in the core are open areas of research.  Some current simulations have more accurate treatments of these issues than we have used here.  However, obtaining reliable cross sections for neutrino scattering with nuclear matter via Standard Model interactions remains problematic, in part because of the difficulty inherent in modeling nuclear matter. The standard neutrino interactions are treated with approximations, just as we have treated non-standard interactions with approximations.   Our approximations are not a result of mysterious properties of FCNCs, but rather stem from uncertainties in matter at high density.  The biggest uncertainty in our calculation does not come from the computational approximations in our model, but rather from this lack of knowledge. We point out these issues to differentiate {\em physical} approximations from {\em computational} approximations. When accurate and reliable EOS and compositions in nuclear matter in the core are available, standard and non-standard types of neutrino scattering can be correctly accounted for.   

A full supernova simulation, with neutrino transport and hydrodynamics, is needed to properly show all the effects of neutrino FCNCs. There are many pieces of known physics that are being tested for relevance in explaining supernova explosions \cite{open}. The supernova model cannot be used as a means of discovering or constraining new physics until known physics has been included and tested in simulations. Such simulations can treat neutrino trapping more realistically than we have done. By keeping track of neutrino distributions, such a simulation could handle the issue, discussed in Sec. \ref{dyefcnc}, of mu and tau neutrinos changing flavor and filling holes in the $\nu_e$ sea before electron captures can occur.  There is also a neutrino FCNC interaction with electrons \cite{afg}. This additional opacity source for neutrino flavor changing could be modeled easily in a full simulation. For all of these reasons, a better result for the reduction in $Y_e$ could be obtained.  More sophisticated simulations also may reveal the fate of the shock, as well as changes to the thermodynamic profile of the core. We used a constant collapse rate in our simulation, but in fact the pressure changes resulting from a continually decreasing $Y_e$ would cause a non-uniform collapse rate.  A full simulation would be able to follow the actual rate of collapse, and any consequences of a non-uniform collapse rate. Finally, a full simulation would provide neutrino specta which could reveal some signature of FCNCs in a supernova signal.  The work presented in this paper will serve as a guide to preparing such a full simulation.

Our results cannot be construed as either favoring or eliminating the existence of FCNCs.  However, they do show that including FCNCs in the current supernova model could cause major changes to the model and its predictions.  It is possible that data from a supernova signal could be used to constrain new physics such as FCNCs. On the other hand, new physics, such as what may be discovered at the LHC, might be required for successful explanation of supernovae. 

\section{Acknowledgements} 
We would like to thank A. Friedland, B. Grinstein, J. Hidaka, and S. Reddy for useful discussions. This work was funded in part by NSF Grant PHY-04-000359 at UCSD and the Terascale Supernova Initiative (TSI) collaboration's DOE  SciDAC grant at UCSD.

\appendix
\section{One Zone Collapse Physics} \label{codeap}
\subsection{Equation of State}
In the liquid drop model we can express the energy of a single nucleus as a sum of bulk, surface, and Coulomb terms,
\begin{equation}
W_N(Y_e,\rho_N,V_N,u) = W_{\rm bulk} + W_{\rm surf} A^{2/3} + W_{\rm coul} A^{5/3}.
\end{equation} 
Here, $V_N$ is the nuclear volume, $u$ is the fraction of the total volume occupied by nuclei, $A$ is the nuclear mass number, and 
$W_{\rm surf}$ and $W_{\rm coul}$ are coefficients of the surface and Coulomb energies, respectively.  Defining $\rho_N$ as the density inside nuclei, we have  
$A=\rho_N V_N$ and $u=\rho/\rho_N$.  The coefficients $W_{\rm surf}$ and $W_{\rm coul}$ are each functions of $Y_e,\rho_N,V_N,u$, and the number density of neutrons, 
$n_n$. We follow Ref.s \cite{bbal,f82} here and take these to be (in MeV)
\begin{eqnarray}
W_{\rm surf} &\approx& 290 \left( \frac{Y_e}{1-X_n} \right )^2\left [ 1- \left( \frac{Y_e}{1-X_n} \right ) \right ]^2, \\
W_{\rm coul}&\approx& 0.75 \left( \frac{Y_e}{1-X_n} \right )^2(1 - 0.236 \rho_{12}^{1/2} + 0.00194\rho_{12}),
\end{eqnarray}
where $\rho_{12} = \rho/(10^{12}{\rm g}/{\rm cm}^3)$.
The value for the mean nuclear mass is found from the free energy minimization condition, $W_{\rm surf} = 2 W_{\rm coul}$. The mean nuclear mass is
\begin{equation} \label{nucmass}
A \approx 194 \left( \frac{Y_e}{1-X_n} \right )^2 (1 - 0.236 \rho_{12}^{1/2})^{-1}.
\end{equation}
We follow Refs. \cite{bbal,f82} and take the kinetic chemical potential (i.e., without rest mass) for neutrons to be (in MeV)
\begin{eqnarray}
\mu_n &\approx& - 16 + 125\left ( 0.5 - \yep \right ) \nonumber \\
&-& 125 \left ( 0.5 -\yep \right )^2 - \left ( \frac{W_{\rm surf}}{2 A^{1/3}}\right )\frac{3-7\yeplong}{1-\yeplong}.
\end{eqnarray}
In this expression we have neglected an additional term, $-[W_{\rm surf}u^{1/3}(1-u^{2/3})]/[4 A^{1/2}(1-3u^{1/3}/2 + u/2)]$. This is justified when $u$ is small.  Neglect of this term will cause some inaccuracy at the highest densities shown in our tables. The neutron-proton kinetic chemical potential difference is (in MeV)
\begin{eqnarray}
\hat{\mu} &=& 250 \left (  0.5 - \yep \right ) \nonumber \\
&-& W_{\rm surf} A^{-1/3} \left[  \yep^{-1} +2\yep^{-1}\frac{1-2\yeplong}{1-\yeplong} \right ].
\end{eqnarray}
The mass fraction of free neutrons in the {\it dilute} limit is
\begin{equation}
X_n \approx 79 \frac{T^{3/2}}{\rho_{10}}e^{\mu_n/T},
\end{equation}
where $\rho_{10}=\rho/(10^{10}{\rm g}/{\rm cm}^3)$. At very high densities neutron degeneracy becomes important and this expression will be inadequate. 
Likewise, in the dilute limit the free proton mass fraction is
\beq
X_p \approx X_n e^{\hat{\mu}/T}.
\eeq

\subsection{Collapse Rate}
Following Refs. \cite{bbal} and \cite{f82} we choose a collapse rate which is a fraction of the free fall rate. The free fall rate for a core with mass $M$ interior to a radius $R$ is
\beq
-\frac{\dot{R}}{R}=\left ( \frac{2G M} {R^3} \right )^{1/2},
\eeq
where $G$ is Newton's Constant.  With $\rho = M/(4\pi R^3/3)$ we have
\beq
\frac {d {\rm ln} R}{ dt} = -\frac{1}{3}\frac{d {\rm ln} \rho}{dt},
\eeq
and so
\beq
\frac{d {\rm ln} \rho}{dt} \approx (224 {\rm s}^{-1})(\rho_{10})^{1/2}.
\eeq
However, as discussed in Ref. \cite{f82}, the actual collapse rate is smaller than this. This is because degenerate electrons provide pressure which slows the collapse.  Hence, we again follow Ref. \cite{f82} and take
\beq
\frac{d {\rm ln} \rho}{dt} \approx (100 {\rm s}^{-1})(\rho_{10})^{1/2} \label{collapserate}
\eeq  
for the collapse rate.  This collapse rate was chosen in Ref. \cite{bbal} to represent the rate found from then existing numerical simulations. Contemporary simulaions \cite{b&y,m&b,jkr}
give similar collapse rates.  The gross features of collapse are set by the low entropy conditions and the rate of electron capture on heavy nuclei (a quantity dominated by the high electron degeneracy).

\subsection{Entropy}
The change in entropy per baryon with density, derived in \cite{f82}, is adopted here:
\begin{eqnarray}
T \left ( \frac {dS}{d\rho} \right ) & = & \left ( \frac {-dY_e}{d\rho} \right )
 \Bigg \{  
[ \mu_e - \hat{\mu} - \delta m_{np} - ( \lambda^\nu_{\rm fp}/ \lambda_{\rm fp} ) ] 
\frac{ (1 - X_n - X_p) \lambda_H }{ A X_p \lambda_{\rm fp} + (1 - X_n - X_p) \lambda_H } \nonumber \\ 
& + &
[ \mu_e - \hat{\mu} - \delta m_{np} - ( \lambda^\nu_{\rm H}/ \lambda_{\rm H} )] 
\frac{ A X_p \lambda_{fp} }{ A X_p \lambda_{\rm fp} + (1 - X_n - X_p) \lambda_H }  \Bigg \}.
\end{eqnarray}
The neutron-proton mass difference is $\delta m_{np}\approx 1.293$ MeV, and 
$\lambda^\nu_{\rm fp}$ and $\lambda^\nu_{\rm H}$ are neutrino energy loss rates for electron capture on free protons and heavy nuclei, respectively. The total (with rest mass) electron chemical potential $\mu_e$ is given by
\beq
\mu_e \approx 11.1\,{\rm MeV} (\rho_{10} Y_e)^{1/3} .
\eeq
The entropy per baryon (in units of Boltzmann's constant $k$) for the system of nuclei, nucleons, and electrons employed in our one-zone collapse is
\begin{eqnarray}
S &\approx& 
\left [ \frac{\pi^2 Y_e}{\mu_e} + 0.2467(1-X_n)\right ] T \nonumber \\
&+& \left \{ \frac{2.5(1-X_n)}{A} +\frac{(1-X_n)}{A} {\rm ln} \left [ \frac{39.49 A^{5/2} ( T)^{3/2}}{\rho_{10}(1-X_n)}\right ] \right \} \nonumber \\ 
&+& \left[ 2.5 X_n + X_n {\rm ln} \left ( \frac {79.07 (T)^{3/2}}{\rho_{10}X_n} \right )\right ],
\end{eqnarray}
where $T$ and $\mu_e$ are in MeV.

\section{Neutrino Nucleus Scattering in Medium} \label{corap}
The general FCNC Lagrangian is given in Eq. (\ref{fcnclag}). For illustrative purposes, we consider the $\nu-d$ term from this Lagrangian and calculate the FCNC cross section for this channel.  For the sake of generality, we drop subscripts on the parameter $\epsilon$. The differential cross section for flavor changing neutrino-nucleus coherent scattering is then \cite{afg}  
\beq \label{diffcs}
\frac{d\sigma}{d {\rm cos}\theta}= 
\frac{\epsilon^2 G_F^2}{\pi} (2N+Z)^2 E_\nu^2 (1+ {\rm cos} \theta) F^2(q).
\eeq
Here, $E_\nu$ is the incident neutrino energy, $N$ and $Z$ are the numbers of neutrons and protons, respectively, in the nucleus, $\theta$ is the scattering angle, and $q$ is the momentum transfer given by
\beq
q=\sqrt{2} E_\nu (1-{\rm cos}\theta)^{1/2}. \label{q}
\eeq
In Eq. (\ref{diffcs}), $F(q)$ is a form factor. In the calculations done in Ref. \cite{afg}, this form factor was set to unity.

In medium, when the neutrino wavelength is comparable to the separation distance between nuclei, that is $1/E_\nu \gtrsim R_c$, interference from multiple nucleus scattering can be accounted for by modifying the form factor.  This is done by 
subtracting a term from the original form factor \cite{sanjay}, 
\beq
\tilde{F}(q)=F(q) - 3 \frac{ {\rm sin} (q R_c) - (q R_c) {\rm cos} (q R_c)} { (qR_c)^3 }
\eeq
and then replacing $F^2(q)$ by $\tilde{F}^2(q)$ in the differential cross section.
In our case, with $F(q)\approx 1$, we have
\beq
\frac{d\sigma}{d {\rm cos}\theta}= 
\frac{\epsilon^2 G_F^2}{\pi} (2N+Z)^2 E_\nu^2 (1+ {\rm cos} \theta)
\left [  1- 3 \frac{ {\rm sin} (q R_c) - (q R_c) {\rm cos} (q R_c)} { (qR_c)^3 } \right ]^2.
\eeq
The cross section is then given by
\beq  \label{cs}
\sigma = \frac{\epsilon^2 G_F^2}{\pi}(2N+Z)^2 E_\nu^2\, \mathcal{I},
\eeq
where $\mathcal{I}$ is the integral of the kinematical factor (1 +cos$\theta$) multiplied 
by $\tilde F ^2(q)$.  Using Eq. (\ref{q}) to change variables, the integral is  
\beq \label{integral}
\mathcal{I} = \int^{2E_\nu}_0 dq 
\left (  \frac{2q}{E_\nu^2} - \frac{q^3}{2 E_\nu^4} \right )
\left (  1- 3 \frac{ {\rm sin} (q R_c) - (q R_c) {\rm cos} (q R_c)} { (qR_c)^3 } \right )^2.
\eeq
The integral $\mathcal{I}$ is evaluated numerically at each density step using that step's values for $E_\nu$ and $R_c=D_{\rm sep}/2$.  

\begin{figure}
\resizebox{9cm}{!}{\includegraphics{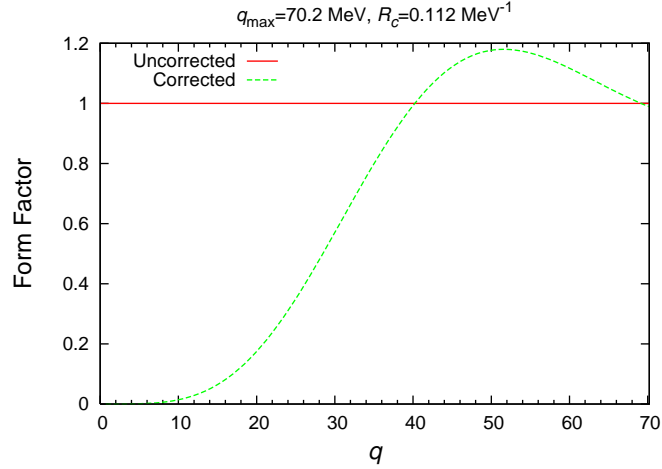}}
\caption{(color online). Corrected form factor as a function of momentum transfer $q$ for core density
 $\rho_{10}= 748$.  At this density $R_c=20.56\,{\rm fm}$, and 
 $E_\nu=35.10\,{\rm MeV}$.}
\label{fffig}
\end{figure}

\begin{figure}
\resizebox{9cm}{!}{\includegraphics{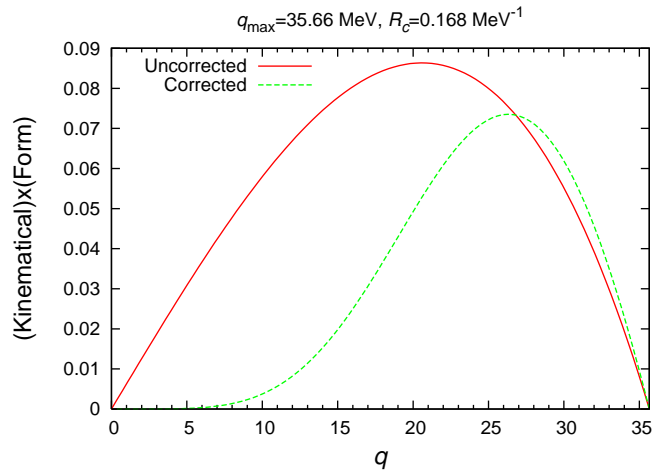}}
\caption{(color online). The solid curve shows kinematical factor multiplied by uncorrected form factor and the dotted curve shows kinematical factor multiplied by the corrected form factor, each plotted as functions of momentum transfer $q$, for core density 
$\rho_{10}= 125$.  At this density $R_c=33.01\,{\rm fm}$, and $E_\nu=17.83\,{\rm MeV}$.}
\label{kinxform}
\end{figure}

To illustrate how the correction for multiple target scattering can change the cross 
section we have included plots of the correction for particular values of core density. In Figure \ref{fffig} we have plotted the corrected form factor when the core is at a density 
$\rho_{10}=748$.  From Table \ref{eps0tab} we see that
$E_\nu=35.10\, {\rm MeV}$ and $R_c=20.56\, {\rm fm} = 0.112\, {\rm MeV}^{-1}$ at this particular density. From Figure \ref{fffig} we clearly see that for small scattering angles (low values of $q$) the interference is destructive and the differential cross section is reduced.  However, we must integrate over $q$. In Figure \ref{kinxform} we have plotted the kinematical factor, $(2q/E_\nu^2 - q^3/2 E_\nu^4)$ multiplied first by the uncorrected form factor $F^2(q)=1$ and then by the corrected form factor $\tilde F^2(q)$.  This plot was made for values of $E_\nu$ and $R_c$ corresponding to density $\rho_{10}=125$.  Table \ref{eps0tab} gives $E_\nu=17.83\,{\rm MeV}$ and 
$R_c= 33.01\,{\rm fm} = 0.168\, {\rm MeV}^{-1}$ at $\rho_{10}=125$.  At this lower density we see that the differential cross is lower over the whole range of $q$.  The interference causes the integrated cross section to be reduced.

\pagebreak

\newpage

\newpage

\newpage

\newpage

\newpage
\begin{table}
\begin{tabular}{|c|c|c|c|c|c|c|c|c|c|c|c|c|c|}\hline 
$\rho_{10}$ & $Y_e$  & $Y_{\nu_e}$ & $Y_{\nu_\mu} + Y_{\nu_\tau}$ & $T\,{\rm (MeV)}$ & $s/k$  &  $X_n$ & $\mu_n\,{\rm (MeV)}$ & $A$ &
$r_{\rm nuc}\,{\rm (fm)}$ & $D_{\rm sep}\,{\rm (fm)}$& $E_{\nu}\,{\rm (MeV)}$ & 
$\lambda_{\nu}\,{\rm (fm)}$\\
\hline

    0.37  &  0.4200  &  0.0000  &  0.0000  &  0.660  &  0.90  &  0.0024  &  -7.10  &   67.  &  4.07  &  387.26  &   0.00  &     inf \\
    0.58  &  0.4200  &  0.0000  &  0.0000  &  0.720  &  0.90  &  0.0041  &  -7.16  &   68.  &  4.08  &  334.17  &   0.00  &     inf \\
    0.91  &  0.4200  &  0.0000  &  0.0000  &  0.782  &  0.90  &  0.0059  &  -7.22  &   68.  &  4.08  &  288.44  &   0.00  &     inf \\
    1.42  &  0.4200  &  0.0000  &  0.0000  &  0.847  &  0.91  &  0.0080  &  -7.29  &   68.  &  4.09  &  249.08  &   0.00  &     inf \\
    2.22  &  0.4198  &  0.0000  &  0.0000  &  0.914  &  0.91  &  0.0101  &  -7.35  &   69.  &  4.10  &  215.22  &   0.00  &     inf \\
    3.47  &  0.4193  &  0.0000  &  0.0000  &  0.984  &  0.91  &  0.0123  &  -7.38  &   70.  &  4.11  &  186.15  &   0.00  &     inf \\
    5.43  &  0.4177  &  0.0000  &  0.0000  &  1.055  &  0.91  &  0.0149  &  -7.34  &   71.  &  4.13  &  161.29  &   0.00  &     inf \\
    8.50  &  0.4143  &  0.0000  &  0.0000  &  1.124  &  0.91  &  0.0185  &  -7.19  &   72.  &  4.17  &  140.15  &   0.00  &     inf \\
   13.30  &  0.4084  &  0.0000  &  0.0000  &  1.190  &  0.92  &  0.0238  &  -6.88  &   75.  &  4.21  &  122.23  &   0.00  &     inf \\
   20.81  &  0.3998  &  0.0000  &  0.0000  &  1.251  &  0.94  &  0.0313  &  -6.42  &   78.  &  4.27  &  107.02  &   0.00  &     inf \\
   32.57  &  0.3884  &  0.0000  &  0.0000  &  1.313  &  0.97  &  0.0425  &  -5.84  &   82.  &  4.34  &   94.10  &   0.00  &     inf \\
   50.96  &  0.3736  &  0.0009  &  0.0000  &  1.375  &  1.01  &  0.0591  &  -5.15  &   87.  &  4.43  &   83.18  &   3.73  &  331.39 \\
   79.74  &  0.3559  &  0.0185  &  0.0000  &  1.372  &  1.01  &  0.0757  &  -4.18  &   94.  &  4.54  &   73.96  &  11.95  &  103.52 \\
  124.77  &  0.3351  &  0.0394  &  0.0000  &  1.351  &  1.01  &  0.0990  &  -3.12  &  103.  &  4.68  &   66.17  &  17.83  &   69.37 \\
  195.24  &  0.3244  &  0.0501  &  0.0000  &  1.430  &  1.01  &  0.1130  &  -2.59  &  111.  &  4.80  &   58.78  &  22.43  &   55.15 \\
  305.51  &  0.3244  &  0.0501  &  0.0000  &  1.622  &  1.01  &  0.1120  &  -2.53  &  119.  &  4.92  &   51.82  &  26.04  &   47.50 \\
  478.07  &  0.3244  &  0.0501  &  0.0000  &  1.834  &  1.01  &  0.1096  &  -2.42  &  130.  &  5.07  &   45.96  &  30.23  &   40.91 \\
  748.07  &  0.3244  &  0.0501  &  0.0000  &  2.064  &  1.01  &  0.1054  &  -2.25  &  146.  &  5.27  &   41.11  &  35.10  &   35.24 \\
 1170.57  &  0.3244  &  0.0501  &  0.0000  &  2.310  &  1.01  &  0.0995  &  -2.00  &  171.  &  5.55  &   37.22  &  40.75  &   30.36 \\
 1831.70  &  0.3244  &  0.0501  &  0.0000  &  2.564  &  1.01  &  0.0917  &  -1.69  &  212.  &  5.96  &   34.34  &  47.31  &   26.15 \\
 2866.21  &  0.3244  &  0.0501  &  0.0000  &  2.821  &  1.01  &  0.0825  &  -1.30  &  292.  &  6.63  &   32.79  &  54.92  &   22.52 \\
 	            
\hline
\end{tabular}
\caption{Sample calculation with no flavor changing coupling, i.e., $\epsilon = 0$. Values are density $\rho_{10}$, electron fraction $Y_e$, mu plus tau neutrino fraction $Y_{\nu_\mu}+Y_{\nu_\tau}$, temperature $T$ in MeV, entropy per baryon $s/k$ in units of Boltzmann's constant, free neutron fraction $X_n$, neutron kinetic chemical potential 
$\mu_n$ in MeV, mean nuclear mass $A$, mean nuclear radius $r_{\rm nuc}$ in fm, separation distance between nuclei $D_{\rm sep}$ in fm, average neutrino energy $E_{\nu}$ in MeV, and average neutrino de Broglie wavelength $\lambda_{\nu}$ in fm. }
\label{eps0tab}
\end{table}

\newpage
\begin{table}
\begin{tabular}{|c|c|c|c|c|c|c|c|c|c|c|c|c|c|}\hline 
$\rho_{10}$ & $Y_e$  & $Y_{\nu_e}$ & $Y_{\nu_\mu} + Y_{\nu_\tau}$ & $T\,{\rm (MeV)}$ & $s/k$  &  $X_n$ & $\mu_n\,{\rm (MeV)}$ & $A$ &
$r_{\rm nuc}\,{\rm (fm)}$ & $D_{\rm sep}\,{\rm (fm)}$& $E_{\nu}\,{\rm (MeV)}$ & 
$\lambda_{\nu}\,{\rm (fm)}$\\
\hline

    0.37  &  0.4200  &  0.0000  &  0.0000  &  0.660  &  0.90  &  0.0024  &  -7.10  &   67.  &  4.07  &  387.26  &   0.00  &     inf \\
    0.58  &  0.4200  &  0.0000  &  0.0000  &  0.720  &  0.90  &  0.0041  &  -7.16  &   68.  &  4.08  &  334.17  &   0.00  &     inf \\
    0.91  &  0.4200  &  0.0000  &  0.0000  &  0.782  &  0.90  &  0.0059  &  -7.22  &   68.  &  4.08  &  288.44  &   0.00  &     inf \\
    1.42  &  0.4200  &  0.0000  &  0.0000  &  0.847  &  0.91  &  0.0080  &  -7.29  &   68.  &  4.09  &  249.08  &   0.00  &     inf \\
    2.22  &  0.4198  &  0.0000  &  0.0000  &  0.914  &  0.91  &  0.0101  &  -7.35  &   69.  &  4.10  &  215.22  &   0.00  &     inf \\
    3.47  &  0.4193  &  0.0000  &  0.0000  &  0.984  &  0.91  &  0.0123  &  -7.38  &   70.  &  4.11  &  186.15  &   0.00  &     inf \\
    5.43  &  0.4177  &  0.0000  &  0.0000  &  1.055  &  0.91  &  0.0149  &  -7.34  &   71.  &  4.13  &  161.29  &   0.00  &     inf \\
    8.50  &  0.4143  &  0.0000  &  0.0000  &  1.124  &  0.91  &  0.0185  &  -7.19  &   72.  &  4.17  &  140.15  &   0.00  &     inf \\
   13.30  &  0.4084  &  0.0000  &  0.0000  &  1.190  &  0.92  &  0.0238  &  -6.88  &   75.  &  4.21  &  122.23  &   0.00  &     inf \\
   20.81  &  0.3998  &  0.0000  &  0.0000  &  1.251  &  0.94  &  0.0313  &  -6.42  &   78.  &  4.27  &  107.02  &   0.00  &     inf \\
   32.57  &  0.3884  &  0.0000  &  0.0000  &  1.313  &  0.97  &  0.0425  &  -5.84  &   82.  &  4.34  &   94.10  &   0.00  &     inf \\
   50.96  &  0.3736  &  0.0009  &  0.0000  &  1.375  &  1.01  &  0.0591  &  -5.15  &   87.  &  4.43  &   83.18  &   3.73  &  331.39 \\
   79.74  &  0.3559  &  0.0185  &  0.0000  &  1.372  &  1.01  &  0.0757  &  -4.18  &   94.  &  4.54  &   73.96  &  11.95  &  103.52 \\
  124.77  &  0.3349  &  0.0393  &  0.0002  &  1.349  &  1.01  &  0.0992  &  -3.11  &  103.  &  4.68  &   66.18  &  17.82  &   69.39 \\
  195.24  &  0.3230  &  0.0501  &  0.0013  &  1.419  &  1.01  &  0.1148  &  -2.53  &  111.  &  4.81  &   58.86  &  22.43  &   55.14 \\
  305.51  &  0.3207  &  0.0501  &  0.0036  &  1.593  &  1.01  &  0.1170  &  -2.38  &  120.  &  4.93  &   52.03  &  26.04  &   47.49 \\
  478.07  &  0.3165  &  0.0501  &  0.0079  &  1.774  &  1.01  &  0.1208  &  -2.08  &  132.  &  5.09  &   46.37  &  30.24  &   40.91 \\
  748.07  &  0.3080  &  0.0501  &  0.0163  &  1.951  &  1.01  &  0.1296  &  -1.56  &  150.  &  5.32  &   41.86  &  35.10  &   35.24 \\
 1170.57  &  0.2906  &  0.0501  &  0.0338  &  2.110  &  1.01  &  0.1528  &  -0.64  &  180.  &  5.65  &   38.66  &  40.75  &   30.35 \\
 1831.70  &  0.2526  &  0.0501  &  0.0717  &  2.269  &  1.01  &  0.2201  &   0.91  &  235.  &  6.17  &   37.36  &  47.31  &   26.14 \\
 2866.21  &  0.2242  &  0.0501  &  0.1001  &  2.612  &  1.01  &  0.2724  &   2.22  &  334.  &  6.94  &   37.06  &  54.93  &   22.52 \\
  	            
\hline
\end{tabular}
\caption{Same as Table \ref{eps0tab} but now with flavor changing coupling $\epsilon = 10^{-3}$.}
\label{eps1-3tab}
\end{table}

\newpage
\begin{table}
\begin{tabular}{|c|c|c|c|c|c|c|c|c|c|c|c|c|c|}\hline 
$\rho_{10}$ & $Y_e$  & $Y_{\nu_e}$ & $Y_{\nu_\mu} + Y_{\nu_\tau}$ & $T\,{\rm (MeV)}$ & $s/k$  &  $X_n$ & $\mu_n\,{\rm (MeV)}$ & $A$ &
$r_{\rm nuc}\,{\rm (fm)}$ & $D_{\rm sep}\,{\rm (fm)}$& $E_{\nu}\,{\rm (MeV)}$ & 
$\lambda_{\nu}\,{\rm (fm)}$\\
\hline

    0.37  &  0.4200  &  0.0000  &  0.0000  &  0.660  &  0.90  &  0.0024  &  -7.10  &   67.  &  4.07  &  387.26  &   0.00  &     inf \\
    0.58  &  0.4200  &  0.0000  &  0.0000  &  0.720  &  0.90  &  0.0041  &  -7.16  &   68.  &  4.08  &  334.17  &   0.00  &     inf \\
    0.91  &  0.4200  &  0.0000  &  0.0000  &  0.782  &  0.90  &  0.0059  &  -7.22  &   68.  &  4.08  &  288.44  &   0.00  &     inf \\
    1.42  &  0.4200  &  0.0000  &  0.0000  &  0.847  &  0.91  &  0.0080  &  -7.29  &   68.  &  4.09  &  249.08  &   0.00  &     inf \\
    2.22  &  0.4198  &  0.0000  &  0.0000  &  0.914  &  0.91  &  0.0101  &  -7.35  &   69.  &  4.10  &  215.22  &   0.00  &     inf \\
    3.47  &  0.4193  &  0.0000  &  0.0000  &  0.984  &  0.91  &  0.0123  &  -7.38  &   70.  &  4.11  &  186.15  &   0.00  &     inf \\
    5.43  &  0.4177  &  0.0000  &  0.0000  &  1.055  &  0.91  &  0.0149  &  -7.34  &   71.  &  4.13  &  161.29  &   0.00  &     inf \\
    8.50  &  0.4143  &  0.0000  &  0.0000  &  1.124  &  0.91  &  0.0185  &  -7.19  &   72.  &  4.17  &  140.15  &   0.00  &     inf \\
   13.30  &  0.4084  &  0.0000  &  0.0000  &  1.190  &  0.92  &  0.0238  &  -6.88  &   75.  &  4.21  &  122.23  &   0.00  &     inf \\
   20.81  &  0.3998  &  0.0000  &  0.0000  &  1.251  &  0.94  &  0.0313  &  -6.42  &   78.  &  4.27  &  107.02  &   0.00  &     inf \\
   32.57  &  0.3884  &  0.0000  &  0.0000  &  1.313  &  0.97  &  0.0425  &  -5.84  &   82.  &  4.34  &   94.10  &   0.00  &     inf \\
   50.96  &  0.3736  &  0.0009  &  0.0000  &  1.375  &  1.01  &  0.0591  &  -5.15  &   87.  &  4.43  &   83.18  &   3.73  &  331.39 \\
   79.74  &  0.3550  &  0.0184  &  0.0010  &  1.365  &  1.01  &  0.0766  &  -4.13  &   94.  &  4.55  &   74.03  &  11.92  &  103.77 \\
  124.77  &  0.3192  &  0.0355  &  0.0197  &  1.215  &  1.01  &  0.1197  &  -2.38  &  106.  &  4.73  &   67.34  &  17.23  &   71.78 \\
  195.24  &  0.2313  &  0.0429  &  0.1003  &  0.811  &  1.01  &  0.3032  &   0.02  &  123.  &  4.97  &   65.94  &  21.30  &   58.08 \\
  305.51  &  0.2265  &  0.0477  &  0.1003  &  1.001  &  1.01  &  0.3138  &   0.19  &  132.  &  5.10  &   58.53  &  25.61  &   48.29 \\
  478.07  &  0.2241  &  0.0500  &  0.1003  &  1.239  &  1.01  &  0.3165  &   0.41  &  145.  &  5.26  &   52.10  &  30.22  &   40.93 \\
  748.07  &  0.2241  &  0.0500  &  0.1003  &  1.528  &  1.01  &  0.3104  &   0.68  &  164.  &  5.48  &   46.58  &  35.09  &   35.25 \\

\hline
\end{tabular}
\caption{Same as Table \ref{eps0tab} but now with flavor changing coupling $\epsilon = 10^{-2}$.}
\label{eps1-2tab}
\end{table}

\end{document}